\documentclass[journal,comsoc,12pt]{IEEEtran}
\usepackage{amsmath,amsfonts}
\usepackage{algorithmic}
\usepackage{algorithm}
\usepackage{array}
\usepackage[caption=false,font=normalsize,labelfont=sf,textfont=sf]{subfig}
\usepackage{textcomp}
\usepackage{stfloats}
\usepackage{url}
\usepackage{verbatim}
\usepackage{graphicx}
\usepackage{cite}
\usepackage{hyperref}
\usepackage{cleveref}
\usepackage{xspace}
\usepackage{graphicx}
\usepackage[inline]{enumitem}

\newcommand{\fw}{0.75}

\begin{document}

\title{A Multi-Stakeholder Perspective on\\Self-Managing Networks}

\author{
	\IEEEauthorblockN{
		Patrick Weber\IEEEauthorrefmark{1}, 
		Artur Sterz\IEEEauthorrefmark{2},
		Bernd Freisleben\IEEEauthorrefmark{2}, 
		Oliver Hinz\IEEEauthorrefmark{1}, 
	}\\
	\IEEEauthorblockA{
		\IEEEauthorrefmark{1}\textit{Chair of Information Systems \& Information Management},
		Goethe University Frankfurt, Germany
	}\\
	\IEEEauthorblockA{
		\IEEEauthorrefmark{2}\textit{Department of Mathematics \& Computer Science},
		University of Marburg, Germany
	}
}

\maketitle

\begin{abstract}
Modern telecommunication networks face an increasing complexity due to the rapidly growing number of networked devices and rising amounts of data.
The literature advocates for self-managing networks as a means to tackle the resulting challenges.
While self-managing networks provide potential solutions to these challenges, current research solely focuses on the perspective of network operators.
However, modern telecommunication networks involve various stakeholders, such as service providers and end users, and necessitate interactions between them.
By transitioning from a single-stakeholder to a \emph{multi}-stakeholder perspective, we address the preferences of all involved parties, acknowledging potential conflicts of interest and constraints like information asymmetries. 
This broader perspective facilitates the development of more effective self-managing networks, significantly enhancing their performance metrics compared to approaches that solely prioritize the concerns of network operators.
\end{abstract}
\begin{IEEEkeywords}
self-managing networks, multi-stakeholder, information asymmetry, preferences
\end{IEEEkeywords}
\section{Introduction}
Modern telecommunication networks face an increasing complexity, e.g., due to the rapidly growing number of devices accessing these networks or the rising amount of data transmitted and processed.
According to an Ericsson report, the total global mobile network traffic reached 160 Exabyte per month in 2023.
For 2029, Ericsson expects a total global mobile network traffic of 403 Exabyte per month, an increase of more than 150\%~\cite{jonsson2024ericsson}.
To cope with the challenges this vast increase of network traffic will introduce, the literature proposed various forms of network management, such as active network management, policy-based network management, automatic network management, or zero-touch network management, which we summarize as \emph{self-managing networks}~\cite{coronado2022zero}.
Currently, the deployment of self-managing networks is in its early stages, with many research questions pending.
However, first examples of self-managing networks include Automatic Neighbor Relation (ANR) of LTE and later 3GPP networks or Cisco's Zero-Touch Provisioning feature for large enterprise networks.

While self-managing networks have the potential to provide solutions to the challenges sketched above, research in the area of self-managing networks currently focuses on improving technical aspects of future networks only from the perspective of the network operator (NO), e.g., increasing bandwidth or reducing latency.
However, in modern telecommunication networks, various stakeholders own, use, and offer distributed communication, computation, storage, and sensing resources.
Services and technologies in these networks, such as fog computing facilities, mobile crowdsensing, and \emph{NextG}, involve interactions between these stakeholders.
Other research areas of modern telecommunication networks, e.g., in the area of edge computing~\cite{cao2020edge} or data platforms~\cite{ahokangas2021platform}, showed that considering multiple stakeholders greatly improves their corresponding goals or achieves better results.

We advocate transitioning from a single stakeholder to a \emph{multi}-stakeholder perspective on self-managing networks to encompass all stakeholders' preferences, considering potentially conflicting interests and constraints like information asymmetries.
Expanding the perspective from a single stakeholder to a multi-stakeholder perspective introduces new challenges.
Stakeholders may have different goals and conflicting preferences, and they may possess different types of information, hindering finding optimal solutions for given problems. However, they may also have beneficial interdependencies, such as spillovers making win-win-situations possible.

In this article, we discuss important aspects of a multi-stakeholder perspective on self-managing networks for the first time and propose solutions to researchers and practitioners of self-managed networks.
We first introduce a system model of self-managing networks, highlighting the shortcomings of the current, single stakeholder model described in the literature (\Cref{sec:scope}).
Afterwards, we discuss these shortcomings and potential solutions.
This includes the definition and identification of stakeholders in self-managing networks (\Cref{sec:stakeholders}).
Each of the stakeholders has different goals and preferences, which require assessment and inclusion in a specific stakeholder's utility function.
Other important aspects include conflicting interests between different stakeholders and information asymmetries.
In \Cref{sec:challenges}, we discuss how to address these issues.
Finally, in \Cref{sec:apply} we present three case studies where we highlight the benefits of a multi-stakeholder perspective on self-managing networks.

\section{Self-Managing Networks as a Continuous Cycle}
\label{sec:scope}
\begin{figure*}
  \centering
  \includegraphics[width=\fw\textwidth]{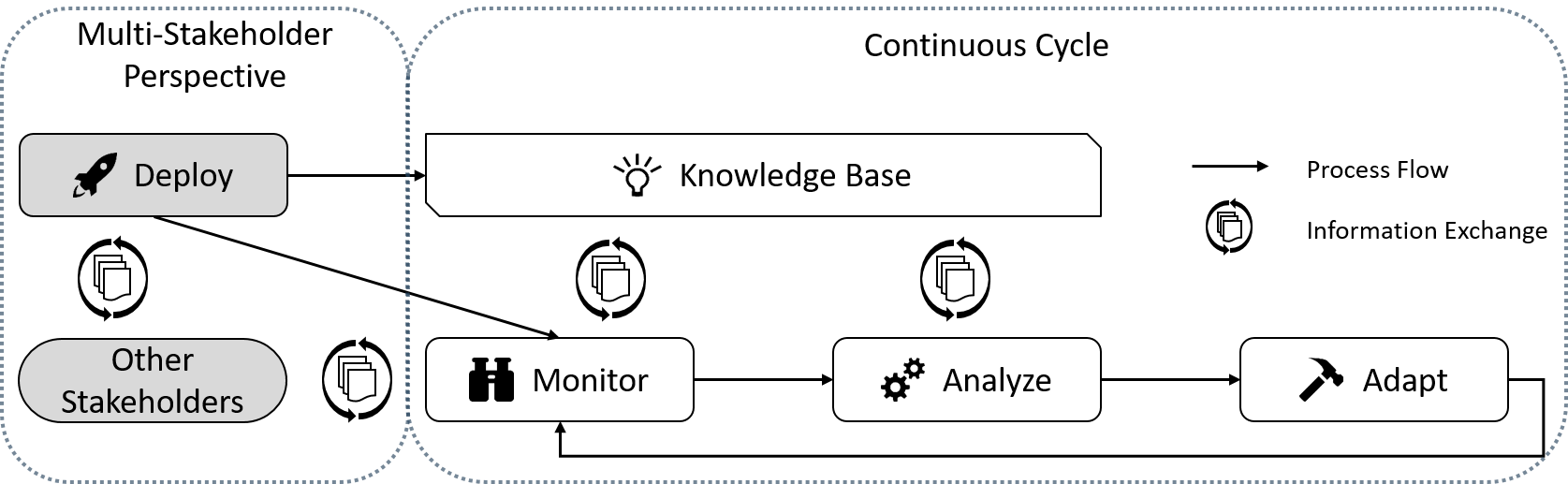}
\caption{An overview of our proposed continuous cycle}
\label{fig:cycle}
\end{figure*}

Operating a self-managing network follows a continuous cycle~\cite{coronado2022zero} (\Cref{fig:cycle}), which begins with continuously \emph{monitoring} the system.
This step encompasses monitoring technical factors of the network, such as the achieved quality of service (QoS), but also special events such as errors or network outages.
The monitored data necessitates continuous \emph{analysis}, a consideration crucial for designers of self-managing networks.
Finally, the information gained from continuous monitoring and analysis might suggest that the performance of the self-managing network may not be optimal or degrading over time due to outages.
Therefore, a self-managing network needs the ability to \emph{adapt} to a new situation.
The information gathered through monitoring and analysis contributes to a common \emph{knowledge base}.
The knowledge base contains all relevant, monitored, and analyzed data and continuous updates throughout the process.
The right part of \Cref{fig:cycle} visualizes this basic process.

We extend this approach described in the literature by an initial \emph{deploy} step, as shown in the left part of \Cref{fig:cycle}.
During this step, researchers and practitioners need to gather relevant information about the given scenario. We argue that knowing and understanding \emph{other stakeholders} for a given scenario is essential.
Each stakeholder in self-managing networks has its own preferences and utilities that must be considered, since they fundamentally influence the performance, but are also influenced by the perceived QoS of a network.
Therefore, the first step is to identify relevant stakeholders for a given scenario, including the stakeholders' preferences and utilities.
This information contributes to a knowledge base future decision making rests on.
Finally, the addition of stakeholders into the continuous cycle also requires adjustments to the process. Properties regarding the stakeholders themselves require continuous monitoring, as their preferences might change over time.

\section{Stakeholders in Self-managing Networks}
\label{sec:stakeholders}

\begin{figure*}
  \centering
  \includegraphics[width=\fw\textwidth]{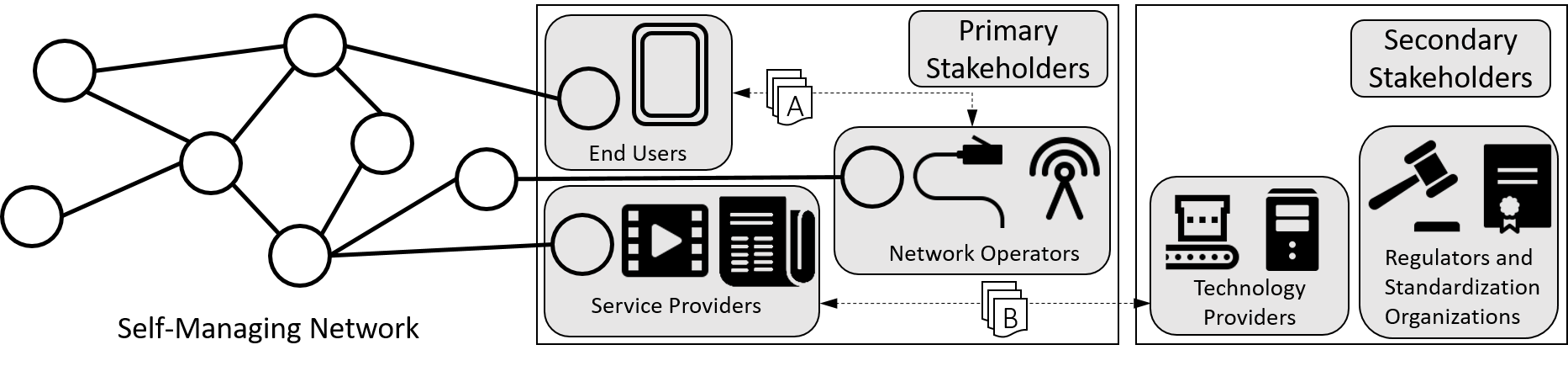}
\caption{An overview of the stakeholders in modern telecommunication networks}
\label{fig:stakeholder_overview}
\end{figure*}

\subsection{Types of Stakeholders}
\label{sub:stakeholders}
The process of designing and developing a self-managing network entails the crucial step of identifying pertinent stakeholders specific to the scenario and problem at hand.
Stakeholders are autonomous decision makers capable of influencing others or being influenced by decisions made by other stakeholders~\cite{Harrison2010managing}.
Additionally, each stakeholder maximizes its utility function, encompassing technical and potentially economic attributes, and factoring in cost-benefit analyses along with individual stakeholder preferences.
This approach ensures a comprehensive consideration of diverse factors in optimizing the functionality and performance of a self-managing network.
We identified the following stakeholders in the field of self-managing networks~\cite{coronado2022zero}, forming two categories:

\subsubsection{Primary Stakeholders}
Primary stakeholders are directly involved in the self-managing network and its outcome (\Cref{fig:stakeholder_overview}).
They either provide resources or services to or through the network, or use them via the network.

\paragraph{Network Operators}
They are stakeholders responsible for operating a self-managing network, e.g., an Internet Service Provider like T-Mobile operating their cellular network or the IT department of a company operating an enterprise network.
NOs play a critical role in defining high-level objectives and goals achievable through the use of self-managing networks.
They design, implement, configure, and maintain the network, its infrastructure and solutions for self-managing networks, taking into account the identified objectives.
The main objective of NOs is ensuring a high QoS in their networks and alignment with business objectives, such as reducing network operation cost.

\paragraph{Service Providers}
They offer various network-based services, services that use network connectivity or services that other stakeholders use over the network.
For example, Video-on-Demand SPs such as Netflix use T-Mobile's network to deliver their content to end users or the accounting department uses a company's network for their accounting service.
SPs' main objective is maintaining a high QoS to their end users adhering to existing, financial constraints.

\paragraph{End Users}
They are the ultimate recipients of network services. Their experience and feedback influences the strategy, development, and optimization of self-managing networks.
In the mentioned examples, end users are either the company's accountants or recipients of Netflix' Video-on-Demand service.
End users benefit from improved network performance and reliability through self-managing networks.

\subsubsection{Secondary Stakeholders}
These are not directly involved, but influence or are affected by self-managing networks~(\Cref{fig:stakeholder_overview}).
They neither provide nor use resources or services, but define conditions, develop novel technology or enforce legal frameworks.
Below, we do not consider their role in self-managing networks any further, since this exceeds the scope of this article. 

\paragraph{Technology Providers}
These include companies like Cisco or researchers from universities who research, develop, produce, and supply the hardware and software solutions required by NOs.

\paragraph{Regulators and Standardization Organizations}
Regulators like the Federal Communications Commission set regulations that guide and influence the legal and ethical framework of researchers and practitioners. 
Standardization organizations like 3GPP develop and enforce standards that ensure interoperability and consistency between different solutions for self-managing networks.

\subsection{Stakeholder Identification}
Researchers and practitioners of self-managing networks must first identify the stakeholders relevant for a given scenario or the application under consideration.
To illustrate stakeholder identification, consider an exemplary scenario of a corporate network where the network needs to self-configure new employee devices and then self-optimize to make the best use of available resources.
Depending on the focal problem, different stakeholders emerge.
For example, for the self-configuration of the network, the relevant stakeholders could be an employee in the role of an end user on the one hand, and the NO of the corporate network on the other hand.
The problem analyzed in this scenario could be the provisioning of the employee's device, where the employee's goal is to access the company's network resources with the best possible QoS, while the NO pursues to provide this service simultaneously meeting the company's requirements regarding, e.g., IT security standards or other service level agreements.
However, in the same scenario, but with different problem statements, the stakeholders could be different, e.g., SPs providing an accounting service to end users aiming to achieve a good QoS and increasing its net profit.

In some scenarios, researchers and practitioners of self-managing networks may not be able to distinguish between stakeholders.
Factors such as vertical integration, driven by SPs' significant investments in infrastructure, blur the distinction between SPs and NOs.
This evolving landscape necessitates considering varying levels of abstraction when modeling stakeholders.
While specific companies can be individually modeled with unique utility functions and preferences, researchers and practitioners of self-managing networks may represent more abstract stakeholders by generalizations that capture essential characteristics which multiple stakeholders fulfilling similar roles in the given scenario share.

\subsection{Preferences}
\label{sub:preferences}

\subsubsection{Assessment of Relevant Factors}
First, researchers and practitioners of self-managing networks need to assess the technical and potentially economic factors in a given scenario.
For each stakeholder, researchers and practitioners of self-managing networks need to identify those parameters of relevance to the respective stakeholder and consider each stakeholder individually.
For example, for the ANR example mentioned above, relevant technical factors include the end users' mobility, transmission range, and transfer speeds.

\subsubsection{Translation of Terminology}
Given that stakeholders may lack expertise in all domains (e.g., end users may lack technical knowledge about self-managing networks), it is essential to translate evaluated factors into comprehensible terms for each stakeholder~\cite{keller2021influence}.
This translation involves tailoring terms and explanations in user studies to suit respondents with varying levels of technical proficiency.
For example, user studies may employ different terminology compared to interviews conducted with technical teams at NOs, who possess a deeper understanding of technical fundamentals.
For user studies, it is ineffective to refer to ANR or request transfer speeds in Bits per second, which would be appropriate for expert NOs.
Following preference elicitation~(see \Cref{subsub:preference}), the obtained results require rephrasing into technical terms.
For example, asking an end user how they rate their transfer speeds requires translation into quantitative, technical terms, such as Bits per second.
This ensures that stakeholders involved in implementing or researching self-managing networks effectively utilize and interpret findings.

\subsubsection{Preference Elicitation}
\label{subsub:preference}

\paragraph{Quantitative Preference Elicitation}
To obtain realistic stakeholder preferences, researchers and practitioners of self-managing networks can employ quantitative preference elicitation methods like experiments.
Experiments entail creating a predefined and consistent environment where participants engage in tasks.
By systematically altering specific parameters, researchers and practitioners of self-managing networks can monitor how participants' behavior changes, allowing for the exploration of interrelationships among various technical factors.
These methods facilitate the analysis of stakeholder preferences, the assessment of the significance of technical factors, and the estimation of adoption rates for technologies.
As an example, a NO could ask thousands of end users about different technical aspects of their network, such as transfer speeds or general QoS in non-technical terms, and then use this data to understand how different handover algorithms affect the end users.

\paragraph{Qualitative Preference Elicitation}
These methods allow gathering non-numerical data to decipher how technical factors reflect onto the stakeholders utility functions.
This approach enables the exploration of stakeholder preferences when the sample size is inadequate for quantitative methods.
Implementing qualitative preference elicitation involves, e.g., using surveys with open-ended questions, allowing for the collection of data on aspects such as use cases, implementations, technical requirements, operational costs, and risk tolerance without rigid constraints.
Afterwards, researchers and practitioners of self-managing networks can analyze this data set, detect recurring patterns, categorize them, and empirically explain stakeholder behavior.
For example, when a NO must design a handover algorithm in ANR, the NO could equip participants with prepared mobile devices that log achieved transfer speeds using different handover algorithms and ask the participants afterwards about how they rate the network's quality.

\subsection{Utility Functions}
\label{sub:utilities}
When stakeholders develop or utilize new technologies, they face critical decision-making processes.
For example, as part of the ANR, a cell tower in a self-managing cellular network needs to select the best neighboring cell tower for vertical handovers due to a mobile device's mobility.
Technical aspects such as network coverage and service availability, but also the other stakeholders' preferences influence such decisions.
Additionally, researchers and practitioners of self-managing networks might also consider economic factors, such as operational costs.
The interdependency of these factors contributes to the complexity of decision-making, given the multitude of potential alternatives.
To address this complexity, multi-criteria decision making research offers various methodologies~\cite{aruldoss2013survey}.
Our proposed approach involves constructing a unified utility function by assigning weights to each factor and adjusting them in accordance with stakeholder preferences.

\section{Challenges and Opportunities}
\label{sec:challenges}
A multi-stakeholder perspective on self-managing networks presents challenges related to stakeholder interactions.
Each stakeholder possesses distinct preferences and objectives possibly conflicting with those of other stakeholders.
Contrarily, interactions among stakeholders, including efforts to reduce information asymmetry, can lead to mutually beneficial improvements.

\subsection{Conflicts of Interests}
In multi-stakeholder scenarios, conflicts of interest among stakeholders are common.
A common example is the conflicting interest of end users and a NO running a mobile network.
While end users desire the highest possible QoS (e.g., high bandwidth and low latency) for the lowest price possible, the NO wants to maximize its profits while providing a service accepted by end users.
To solve these conflicts, stakeholders' intentions play a key role to determine whether they seek a collaborative solution and engage in cooperative problem-solving or opt for non-cooperation and foster competitive relationships.
In cases of non-cooperation, additional costs arise for all participants.
These may stem from the necessity of involving a neutral third party, such as a broker, to facilitate a resolution, or from decentralized approaches possibly not yielding optimal results.
However, there are methods from game theory such as the Nash Bargaining Solution (NBS) or Double Auction games to optimize strategies in scenarios with multiple selfish stakeholders~\cite{coronado2022zero}.

\subsection{Information Asymmetries}
Information asymmetries refer to situations in which one stakeholder has more or better information that is not shared with other stakeholders~\cite{dawson2010information}.
For example, only the SP knows about the usage of a particular service, and only the NO knows the network's actual operational costs.
Stakeholders may be committed not to share information due to economic or technical reasons.
For example, the NO could hide the true cost structure of the network operation from the SP in order to achieve a higher profit margin for the use of the network.
There are four methods to reduce information asymmetry~\cite{dawson2010information}:
Signaling and information exchange, where the better informed stakeholder shares information either before or after the agreement; screening and monitoring, where the less informed stakeholder observes behavior, actions or third-party information before or after the agreement.
Machine learning approaches enable stakeholders to optimize their utility in the given scenario by learning from past decisions, interactions, or screened and monitored information.
For example, since the NO might not know the mobility of its end users, it can monitor the end users, learn their movement patterns and thus improve the vertical handover of ANRs using machine learning.

\subsection{Mutually Beneficial Interdependencies}
Despite conflicting interests and information asymmetries, stakeholders in self-managing networks may also experience beneficial interactions.
For instance, if a NO enhances the handover capabilities of its mobile network using an ANR protocol designed with preferences of end users of a particular service, all other end users might also experience better handover capabilities.
Simultaneously, the introduction of new capabilities can attract new end users, opening up novel opportunities for the NO and paving the way for future enhancements.

\section{Bringing it Together}
\label{sec:apply}
We present three case studies, where we applied the proposed multi-stakeholder perspective on self-managing networks: a \emph{device-to-device (D2D) data transfer application}~\cite{sterz2020reactifi}, an application providing an \emph{energy-optimal multi-hop network for Internet-of-Things (IoT) scenarios}~\cite{sterz2023energy}, and a \emph{placement application for edge computing scenarios}~\cite{sterz2022multi}.

In the first case study on a \emph{D2D application}, the objective is implementing an algorithm for optimizing data throughput between two mobile devices within a local self-managing network tailored for data transfer applications such as Apple AirDrop or Google Quick Share.
The stakeholders involved are a NO that ensures that the end users accessing the network at the same time receive a satisfactory level of service, and two end users desiring to exchange data as fast as possible.
The NO's utility function contains technical parameters like allocated bandwidth per end user, while end users’ utility functions contain transfer speed and mobility.
To enhance data transfer speed within the local network, we employ the Wi-Fi Tunneled Direct Link Setup (TDLS) technology, introduced by IEEE 802.11z, to seamlessly transition between Access Point (AP) and peer-to-peer (P2P) communication modes based on Received Signal Strength Indicator (RSSI) thresholds as an estimate of the distance between the two end users.
Here, information asymmetry is not an issue, since the stakeholders themselves can sense all required information.
By dynamically switching modes based on RSSI levels, i.e., using the AP mode when the two end users are far apart and TDLS is in close proximity, we achieve optimal data transfer rates while balancing data transmission range and speed capabilities.
This dynamic approach maximizes end users' utility by increasing the average thoughput by up to 230\% (from 12 MBit/s without switching from AP mode to P2P compared to 40 MBit/s when switching to P2P mode in close proximity).
Furthermore, it optimizes the NO's utility by effectively managing Wi-Fi capacity between AP and P2P modes to utilize available Wi-Fi spectrum and bandwidth efficiently.
This result shows that using the proposed multi-stakeholder perspective on self-managing networks achieves better results for both the NO as well as the end users.

In our case study of an \emph{energy-optimal IoT network}, the focus lies in establishing an energy-efficient network for transmitting data from a single source to all other nodes while preserving energy.
The stakeholders involved in this scenario are the NO and a SP, where the NO aims to minimize energy consumption to prolong the network's operational lifespan, in contrast to the SP's objective of fast data dissemination.
The conflicting goals of these stakeholders necessitate the consideration of energy consumption and QoS in the NO's utility function, whereas the SP's utility focuses solely on data transmission aspects.
To address this challenge, we designed a network protocol to optimize energy consumption by constructing an energy-efficient spanning tree, leveraging a game-theoretic approximation to overcome the NP-hard nature of the spanning tree construction problem.
Through simulation-based scalability and real-world IoT testbed evaluations, we showed that the proposed multi-stakeholder perspective enhances energy efficiency for data transmission of up to an average of 70\% within the self-managing network.

The preceding applications highlighted how adopting a multi-stakeholder perspective enhances technical aspects within self-managing networks during operation.
In contrast, the third case study on \emph{edge placement application} focuses on optimizing economic factors at the deployment stage of an edge-computing use case, emphasizing cost reduction for the NO and SP while maintaining or improving QoS for end users.
The use of edge computing, particularly cloudlets, offers cost saving opportunities by strategically placing services on highly utilized cloudlets, thus minimizing data transfer across the NO's network.
The stakeholders' utility functions include operational costs, costs for processing on cloudlets, for data transfer, for hardware provisioning, and for maintenance.
Some of this information, however, is private to the stakeholders, necessitating an agreement on service deployment locations and payment terms under information asymmetries.
To address information asymmetries, we apply the game-theoretic NBS to negotiate cloudlet usage, with stakeholders utilizing publicly available data to further optimize cost reductions in service placement.
Using this multi-stakeholder approach, the NO can reduce its cost for operating the cloudletinfrastructure up to an average of 47\%.
The SP can achieve a cost reduction of an average of about 44\% by placing its service using the multi-stakeholder approach compared to a selfish approach.
This case study illustrates the advantages of a multi-stakeholder perspective in achieving better cost efficiencies for both parties.
\section{Conclusion and Outlook}
\label{sec:conclusion}

We demonstrated the importance of a \emph{multi}-stakeholder perspective on self-managing networks by highlighting its potential for enhancing efficiency, effectiveness, and overall performance of modern telecommunication networks.
We discussed important aspects of multiple stakeholders, their preferences and utility functions and presented three case studies to illustrate the benefits of embracing a multi-stakeholder perspective on self-managing networks.

Future research areas include investigating strategies to foster collaboration among stakeholders with potentially conflicting interests and resolving such conflicts, further enhancing the effectiveness of self-managing networks.
Given the significance of information asymmetries, developing technologies to mitigate such disparities and improve transparency potentially lead to more informed decision-making processes.
Finally, more practical implementations to prove the efficacy of multi-stakeholder approaches in real-world scenarios may provide further valuable empirical evidence.
\section{Acknowledgement}
This work has been funded by the Deutsche Forschungsgemeinschaft (DFG, German Research Foundation) – Project-ID 210487104 - SFB 1053.

\bibliographystyle{IEEEtran}
\bibliography{literature}

\begin{thebibliography}{10}
\providecommand{\url}[1]{#1}
\csname url@samestyle\endcsname
\providecommand{\newblock}{\relax}
\providecommand{\bibinfo}[2]{#2}
\providecommand{\BIBentrySTDinterwordspacing}{\spaceskip=0pt\relax}
\providecommand{\BIBentryALTinterwordstretchfactor}{4}
\providecommand{\BIBentryALTinterwordspacing}{\spaceskip=\fontdimen2\font plus
\BIBentryALTinterwordstretchfactor\fontdimen3\font minus
  \fontdimen4\font\relax}
\providecommand{\BIBforeignlanguage}[2]{{%
\expandafter\ifx\csname l@#1\endcsname\relax
\typeout{** WARNING: IEEEtran.bst: No hyphenation pattern has been}%
\typeout{** loaded for the language `#1'. Using the pattern for}%
\typeout{** the default language instead.}%
\else
\language=\csname l@#1\endcsname
\fi
#2}}
\providecommand{\BIBdecl}{\relax}
\BIBdecl

\bibitem{jonsson2024ericsson}
\BIBentryALTinterwordspacing
P.~Jonsson, P.~Cerwall, A.~Lundvall, M.~Ekstrand, D.~von Koch, F.~Fornstad,
  M.~Arvedson, G.~Blennerud, V.~Chen, L.~Elénius, P.~Lindberg, P.~Linder,
  T.~Lodhi, J.~Travers, J.~Yazlle, and F.~Jejdling, \emph{Ericsson Mobility
  Report Business Review 2024}.\hskip 1em plus 0.5em minus 0.4em\relax
  Stockholm: Telefonaktiebolaget LM Ericsson, Mar 2024. [Online]. Available:
  \url{https://www.ericsson.com/4912e3/assets/local/reports-papers/mobility-report/documents/2024br/emr-business-review-2024.pdf}
\BIBentrySTDinterwordspacing

\bibitem{coronado2022zero}
E.~Coronado, R.~Behravesh, T.~Subramanya, A.~Fernandez-Fernandez, M.~S.
  Siddiqui, X.~Costa-P{\'e}rez, and R.~Riggio, ``Zero touch management: A
  survey of network automation solutions for {5G} and {6G} networks,''
  \emph{IEEE Communications Surveys \& Tutorials}, vol.~24, no.~4, pp.
  2535--2578, 2022.

\bibitem{cao2020edge}
X.~Cao, G.~Tang, D.~Guo, Y.~Li, and W.~Zhang, ``Edge federation: Towards an
  integrated service provisioning model,'' \emph{IEEE/ACM Transactions on
  Networking}, vol.~28, no.~3, pp. 1116--1129, 2020.

\bibitem{ahokangas2021platform}
P.~Ahokangas, M.~Matinmikko-Blue, S.~Yrj{\"o}l{\"a}, and H.~H{\"a}mm{\"a}inen,
  ``Platform configurations for local and private {5G} networks in complex
  industrial multi-stakeholder ecosystems,'' \emph{Telecommunications Policy},
  vol.~45, no.~5, p. 102128, 2021.

\bibitem{Harrison2010managing}
J.~S. Harrison, D.~A. Bosse, and R.~A. Phillips, ``Managing for stakeholders,
  stakeholder utility functions, and competitive advantage,'' \emph{Strategic
  Management Journal}, vol.~31, no.~1, pp. 58--74, 2010.

\bibitem{keller2021influence}
K.~Keller, M.~Ott, O.~Hinz, and A.~Klein, ``Influence of social relationships
  on decisions in device-to-device communication,'' \emph{IEEE Access}, vol.~9,
  pp. 106\,459--106\,475, 2021.

\bibitem{aruldoss2013survey}
M.~Aruldoss, T.~M. Lakshmi, and V.~P. Venkatesan, ``A survey on multi criteria
  decision making methods and its applications,'' \emph{American Journal of
  Information Systems}, vol.~1, no.~1, pp. 31--43, 2013.

\bibitem{dawson2010information}
G.~S. Dawson, R.~T. Watson, and M.-C. Boudreau, ``Information asymmetry in
  information systems consulting: Toward a theory of relationship
  constraints,'' \emph{Journal of Management Information Systems}, vol.~27,
  no.~3, pp. 143--178, 2010.

\bibitem{sterz2020reactifi}
A.~Sterz, M.~Eichholz, R.~Mogk, L.~Baumg{\"a}rtner, P.~Graubner, M.~Hollick,
  M.~Mezini, and B.~Freisleben, ``{ReactiFi}: Reactive programming of {Wi-Fi}
  firmware on mobile devices,'' \emph{The Art, Science, and Engineering of
  Programming}, vol.~5, no.~2, pp. 4--1, 2020.

\bibitem{sterz2023energy}
A.~Sterz, R.~Klose, M.~Sommer, J.~H{\"o}chst, J.~Link, B.~Simon, A.~Klein,
  M.~Hollick, and B.~Freisleben, ``Energy-efficient decentralized broadcasting
  in wireless multi-hop networks,'' \emph{Sensors}, vol.~23, no.~17, p. 7419,
  2023.

\bibitem{sterz2022multi}
A.~Sterz, P.~Felka, B.~Simon, S.~Klos, A.~Klein, O.~Hinz, and B.~Freisleben,
  ``Multi-stakeholder service placement via iterative bargaining with
  incomplete information,'' \emph{IEEE/ACM Transactions on Networking},
  vol.~30, no.~4, pp. 1822--1837, 2022.

\end{thebibliography}

\section{Biographies}
{\small
\textbf{Patrick Weber} (weber@wiwi.uni-frankfurt.de) is pursuing a  Ph.D. in Information Systems at the Chair of Information Systems \& Information Management at Goethe University Frankfurt. His research interests include information asymmetries and explainable artificial intelligence.

\textbf{Artur Sterz} (sterz@informatik.uni-marburg.de) is a Postdoc in the Department of Mathematics \& Computer Science at the Philipps-Universit\"at Marburg.  His research focuses on wireless communication in mobile networks.

\textbf{Bernd Freisleben} [Member] (freisleb@informatik.uni-marburg.de) is a full professor of computer science at the University of Marburg, Germany. His research interests include distributed systems, mobile computing, and networked applications.

\textbf{Oliver Hinz} (ohinz@wiwi.uni-frankfurt.de) is a full professor of Information Systems \& Information Management at Goethe University Frankfurt. He is interested in research at the intersection of technology and markets.

}

\end{document}